# Resonance Fluorescence Spectrum of a Trapped Ion Undergoing Quantum Jumps

V. Bühner and Chr. Tamm

*Physikalisch-Technische Bundesanstalt, Bundesallee 100, D-38116 Braunschweig, Germany*

(March 14, 2000)

## Abstract

We experimentally investigate the resonance fluorescence spectrum of single $^{171}$Yb and $^{172}$Yb ions which are laser cooled to the Lamb-Dicke regime in a radiofrequency trap. While the fluorescence scattering of $^{172}$Yb is continuous, the $^{171}$Yb fluorescence is interrupted by quantum jumps because a nonvanishing rate of spontaneous transitions leads to electron shelving in the metastable hyperfine sublevel $^2D_{3/2}$(F=2). The average duration of the resulting dark periods can be varied by changing the intensity of a repumping laser field. Optical heterodyne detection is employed to analyze the fluorescence spectrum near the Rayleigh elastic scattering peak. It is found that the stochastic modulation of the fluorescence emission by quantum jumps gives rise to a Lorentzian component in the fluorescence spectrum, and that the linewidth of this component varies according to the average duration of the dark fluorescence periods. The experimental observations are in quantitative agreement with theoretical predictions.
PACS numbers(s): 32.80.Pj, 42.50.Lc

Typeset using REVTEX

The resonance fluorescence scattered by a monochromatically excited atomic two-level system generally contains both inelastic and elastic components. Inelastic scattering dominates in the case of saturation and gives rise to spectral components whose linewidth is determined by the radiative damping rate of the atomic dipole [1]. The properties of the inelastic fluorescence spectrum were investigated experimentally for atoms in a thermal beam [2] and for a laser-cooled atomic ion in a radiofrequency trap [3]. The elastic fluorescence scattering spectrum of localized atomic two-level systems which satisfy the Lamb-Dicke condition is dominated by a single monochromatic component at the frequency of the driving field [4]. Recent experiments demonstrate that optical heterodyne detection allows to determine the elastic scattering spectrum of trapped atomic ensembles and of single atomic systems with a resolution that is unaffected by either the natural transition linewidth or the optical excitation linewidth [5,6].

High-resolution spectroscopy of the elastically scattered fluorescence is attractive for a number of investigations on laser cooled trapped atoms since the elastic scattering spectrum yields information on the atomic motion under conditions of continuous optical excitation [5]. In the case of atoms not prepared as two-level systems, the interpretation of the fluorescence spectrum is usually less straightforward since the internal atomic dynamics can introduce additional narrow-bandwidth spectral features [7,8].

This paper reports an experimental investigation on the resonance fluorescence spectrum of a single trapped ion whose fluorescence emission is randomly interrupted by quantum jumps to a metastable level [9]. Our experiment realizes a situation also considered in recent theoretical investigations on the resonance fluorescence spectrum of atomic three-level systems. It is predicted that the random population of a metastable level gives rise to an additional component in the fluorescence power spectrum which appears as a Lorentzian pedestal of the Rayleigh elastic scattering peak. Typically the linewidth of this pedestal is much smaller than the width of the inelastic two-level spectrum [10–12]. A quantitative calculation predicts that width and relative height of the Lorentzian pedestal are uniquely determined by the quantum-jump statistics [11]. Specifically, if $\tau_B$ and $\tau_D$ denote the average lengths of bright and dark fluorescence periods, the full width at half maximum (FWHM) of the pedestal is calculated as $\delta_L = (\tau_B^{-1} + \tau_D^{-1})/\pi$. The expected height ratio of pedestal and elastic scattering peak is $A_L = \pi \delta_R \tau_D^2 / (\tau_B + \tau_D)$ if the fluorescence spectrum is registered with a FWHM resolution bandwidth of $\delta_R$ ($\delta_R \ll \delta_L$).

In our experiment, optical heterodyne detection is used to analyze the resonance fluorescence spectrum of single laser cooled $^{172}$Yb and $^{171}$Yb ions with a frequency resolution of $\delta_R \simeq 1$ Hz. While the fluorescence emission of $^{172}$Yb$^+$ is continuous, the $^{171}$Yb$^+$ fluorescence exhibits dark times due to spontaneous transitions to the F=2 hyperfine sublevel of the metastable $^2D_{3/2}$ state [13]. The natural lifetime of this state is 53 ms [14]. The average length of the dark periods in the $^{171}$Yb$^+$ fluorescence emission can be reduced to less than the natural $^2D_{3/2}$ lifetime by nonresonant optical excitation as described below. We determine the resonance fluorescence spectrum of a trapped $^{171}$Yb ion for several average dark-period durations $\tau_D$ resulting in significantly different expected pedestal linewidths $\delta_L$. The corresponding values of $\tau_D$ and $\tau_B$ are determined in separate measurements where the temporal variation of the fluorescence intensity is registered.

Fig.1 shows a section of the energy level system of $^{171}$Yb$^+$ and the employed optical excitation scheme. For fluorescence scattering and Doppler cooling, a linearly polarized continuous laser field excites the low-frequency wing of the quasi-cycling F=1 ↔ F=0 component of the $^{171}$Yb$^+$ $^2S_{1/2} \leftrightarrow\, ^2P_{1/2}$ resonance transition. The transition is at a wavelength of 370 nm and has a natural linewidth of 23 MHz (FWHM). Optical pumping between the magnetic sublevels of the F=1 ground state is reduced by a suitably oriented static magnetic field. Hyperfine pumping to the F=0 ground state is compensated by excitation of the F=0 ↔ F=1 component of the resonance transition. The F=1 sublevel of the $^2D_{3/2}$ state, which is rapidly populated by spontaneous decay from the $^2P_{1/2}$(F=0) state, is depleted through repumping excitation at 935 nm [15]. The F=2 sublevel of the $^2D_{3/2}$ state is not strongly coupled to the cooling and repumping excitation so that individual transitions to this level give rise to macroscopic fluorescence dark periods. Spontaneous transitions to the $^2D_{3/2}$(F=2) level occur mainly as a result of nonresonant 370-nm excitation to the $^2P_{1/2}$(F=1) level so that the resulting value of $\tau_B$ is essentially determined by the applied cooling laser intensity. The $^2D_{3/2}$(F=2) level is depleted by spontaneous decay and by nonresonant 935-nm excitation to the $^3[3/2]_{1/2}$(F=1) level. The value of $\tau_D$ is thus mainly determined by the 935-nm repumping intensity if the applied intensity is sufficiently large [13]. The quantum jump mechanism described here has no analog in $^{172}$Yb$^+$ since this isotope has no hyperfine structure. If the optical excitation scheme shown in Fig. 1 is applied to $^{172}$Yb$^+$, the fluorescence scattering is continuous [15].

The employed experimental setup is shown in Fig. 2. Yb ions are confined in a cylindrically symmetric radiofrequency trap with a ring electrode diameter of 1.4 mm. The trap is operated to obtain axial and radial secular frequencies in the range of 800 kHz. Under these conditions a trapped Yb ion cooled to the Doppler limit satisfies the Lamb-Dicke condition at 370 nm with respect to the residual motion in the trap pseudopotential (macromotion). The trap setup includes two heatable reservoirs for $^{172}$Yb and $^{171}$Yb and an electron source for ionization. After ion loading, adjustable voltages are applied to electron source, Yb reservoirs, and across the trap endcap electrodes in order to compensate static electric stray fields in the confinement volume in three dimensions. The achieved compensation is sufficient to reduce the amplitude of stray-field induced ion motion at the frequency of the trap field (micromotion) to less than the macromotion amplitude at the Doppler cooling limit [16].

Laser radiation at 370 nm with a power of typically 10 $\mu$W is produced by a frequency-doubled extended-cavity diode laser; hyperfine repumping radiation is generated by modulating of the injection current of the diode laser at $\Delta_S + \Delta_P \simeq 14.7$ GHz [16]. A second extended-cavity diode laser produces 935-nm repumping radiation which is overlapped with the focused 370-nm light in the trap. After passage through the trap, the 370-nm light field is frequency shifted by an acousto-optic modulator in order to obtain the local oscillator field required for heterodyne detection. The intensity of the focused 370-nm cooling radiation is in the range of the saturation intensity of the $^{171}$Yb$^+$ $^2S_{1/2}(F=1) \leftrightarrow ^2P_{1/2}(F=0)$ transition so that the Rayleigh scattering intensity is maximized [17]. The depletion of the

$^2D_{3/2}$(F=1) level and of the F=0 ground state is sufficiently rapid that the fluorescence level observed during bright periods is not significantly reduced by dwell times in these states.

Resonance fluorescence light from the trap is collected over a solid-angle segment of approximately $4\pi/60$ by a diffraction-limited f/1.9 lens system. The collimated fluorescence light is imaged on a photomultiplier tube operated in the counting mode. Maximum detected photon scattering rates are $3.5 \times 10^4$ s$^{-1}$ and $9 \times 10^4$ s$^{-1}$ respectively for single $^{171}$Yb and $^{172}$Yb ions.

In order to observe the temporal variation of the $^{171}$Yb$^+$ fluorescence intensity due to quantum jumps, the local oscillator field used for heterodyne detection is blocked and the photodetection pulse signal is registered after passage through a low-pass filter with 1 ms time constant. Fig. 3 gives an example of a corresponding measurement. The fluctuations of the intensity signal during bright periods are mainly due to photodetection shot noise. During dark periods, the signal level is determined by stray light and photomultiplier dark pulses. The comparison of experimental dark-time histogram and fitted exponential indicates that the observed dark-time statistic does not systematically deviate from the exponential distribution expected in the case of random quantum jumps [9]. Evaluations as shown in Fig. 3 were carried out in order to determine $\tau_D$ and $\tau_B$ for the experimental conditions of the fluorescence spectrum measurements described below. Observed $\tau_D$ values increase from approximately 8 ms to 40 ms if the available 935-nm power of 2 mW is attenuated by a factor of 10; $\tau_B$ is typically in the range of 150 ms and shows variations of up to $\pm 50$ ms as a result of drifts of the available cooling laser power.

In order to observe the fluorescence spectrum through heterodyne detection, a fraction of the frequency-shifted 370-nm cooling laser field is passed through wavefront-matching optics and superimposed with the collimated fluorescence light. Effectively this optical arrangement forms a Mach-Zehnder interferometer whose input beam splitter is replaced by a single trapped ion. The interference of the fluorescence and local oscillator light fields gives rise to a beat component in the frequency spectrum of the photomultiplier output pulse signal at the heterodyne frequency $\nu_H$. The photomultiplier signal is passed through a bandpass filter centered at $\nu_H$ in order to reject broadband noise components. The filtered signal is down-converted in frequency so that its power spectrum can be observed using a fast-Fourier-transform (FFT) analyzer. The power of the local oscillator light field is chosen to be several times larger than that of the collected fluorescence light. Under these conditions, the signal-to-noise ratio (SNR) of the heterodyne signal is essentially independent of the applied local oscillator power and determined by photodetection shot noise [18]. The minimum optical frequency resolution bandwidth achievable in this detection scheme is equal to the resolution with which the spectrum of the heterodyne signal is analyzed. The effective optical frequency resolution $\delta_R$ can however be degraded by uncorrelated low-frequency fluctuations of the optical path lengths of the local oscillator and fluorescence light beams [6]. In order to reduce such perturbations in the present experiment, critical beam paths were shielded against air currents, and the trap setup was decoupled from vibrations of its mechanical support.

Experimentally observed fluorescence spectra are shown in Fig. 4. Each spectrum was averaged over a time of 10 minutes in order to reduce baseline fluctuations due to photodetection shot noise. The Rayleigh scattering peaks of the $^{172}$Yb$^+$ and $^{171}$Yb$^+$ fluorescence spectra are resolved with linewidths not significantly larger than the employed FFT resolution bandwidth of $\delta_R \simeq 1$ Hz. The Rayleigh peak of the $^{172}$Yb$^+$ spectrum is observed with a SNR of approximately $10^3$ (30 dB). Weak sidebands with a frequency offset in the range of 50 Hz visible in Fig. 4(a) are likely caused by ambient acoustic perturbations. The Rayleigh peak signal obtained from $^{171}$Yb$^+$ is significantly smaller than the $^{172}$Yb$^+$ signal so that the resulting SNR is in the range of 18 dB. The comparison of the spectra shown in Fig. 4 indicates that the Rayleigh peaks of the $^{171}$Yb$^+$ spectra exhibit pedestals while the $^{172}$Yb$^+$ spectrum appears without pedestal. As shown in Fig. 4(b-d), the pedestal linewidth of the $^{171}$Yb$^+$ spectrum decreases and the relative pedestal height increases if the average dark-time length $\tau_D$ is increased in the range $\tau_D < \tau_B$.

In Fig. 4(b-d) the experimental $^{171}$Yb$^+$ spectra are compared with Lorentzian pedestal lineshapes calculated according to the expressions quoted above. The experimentally de-termined $\tau_\text{D}$ and $\tau_\text{B}$ values used in the calculation are indicated in the figure caption; an effective resolution bandwidth of $\delta_\text{R} = 1$ Hz was assumed. The comparison of calculated and experimental lineshapes indicates that the shape of the observed $^{171}$Yb$^+$ spectra and the dependence of the lineshape on the quantum-jump statistics are in good agreement with the predictions of Ref. [11]. Within the signal-to-noise ratio of the present experimental data, no systematic deviations between calculated and observed lineshapes are found. The observation that the Rayleigh peak signal of the $^{171}$Yb$^+$ spectrum is significantly weaker than that of $^{172}$Yb$^+$ is in agreement with a calculation extending the treatment of Ref. [7]. It is found that under the realized experimental conditions, optical pumping between the magnetic sublevels of the F=1 ground state of $^{171}$Yb$^+$ significantly enhances the inelastic scattering contribution to the total fluorescence emission while the maximum resonant fluorescence scattering rate is reduced [17]. Both effects reduce the $^{171}$Yb$^+$ heterodyne signal strength relative to $^{172}$Yb$^+$ where ground-state optical pumping does not occur.

In conclusion, we have observed the elastic component of the resonance fluorescence spectrum of a trapped ion under conditions of continuous fluorescence scattering, and for the case that the fluorescence exhibits dark periods due to spontaneous transitions to a metastable level. In quantitative agreement with the calculations of Ref. [11], we find that the random modulation of the fluorescence emission by quantum jumps gives rise to a fluorescence power spectrum characterized by a narrow Lorentzian pedestal centered to the Rayleigh scattering peak. It is worth noting that the observed spectrum is identical in shape to the spectrum that results if the continuous fluorescence of a two-level atom or the output of a monochromatic light source is randomly blocked by a shutter. Hence it appears that the Lorentzian pedestal in the fluorescence spectrum of an atom undergoing quantum jumps is a direct consequence of the random switching of the fluorescence intensity, and not the result of phase or intensity fluctuations of the atomic emission during bright fluorescence periods [11]. This interpretation is corroborated by a calculation which indicates that the Lorentzian pedestal vanishes if the fluorescence spectrum is registered only during bright periods whose duration is much above the average [12]. In order to verify this prediction experimentally, fluorescence spectra and the fluorescence intensity variation due to quantum jumps would have to be registered simultaneously using separate optical detection systems. Other experimental investigations on the fluctuation properties of the fluorescence emission during bright periods could take advantage of extensions of the heterodyne detection scheme described above. In particular, phase sensitive heterodyne detection could be employed in order to monitor the fluctuations in both field quadratures of the emitted fluorescence.

This work was supported by the Deutsche Forschungsgemeinschaft in Project B1 of SFB 407. The technical assistance of D. Griebsch is gratefully acknowledged.

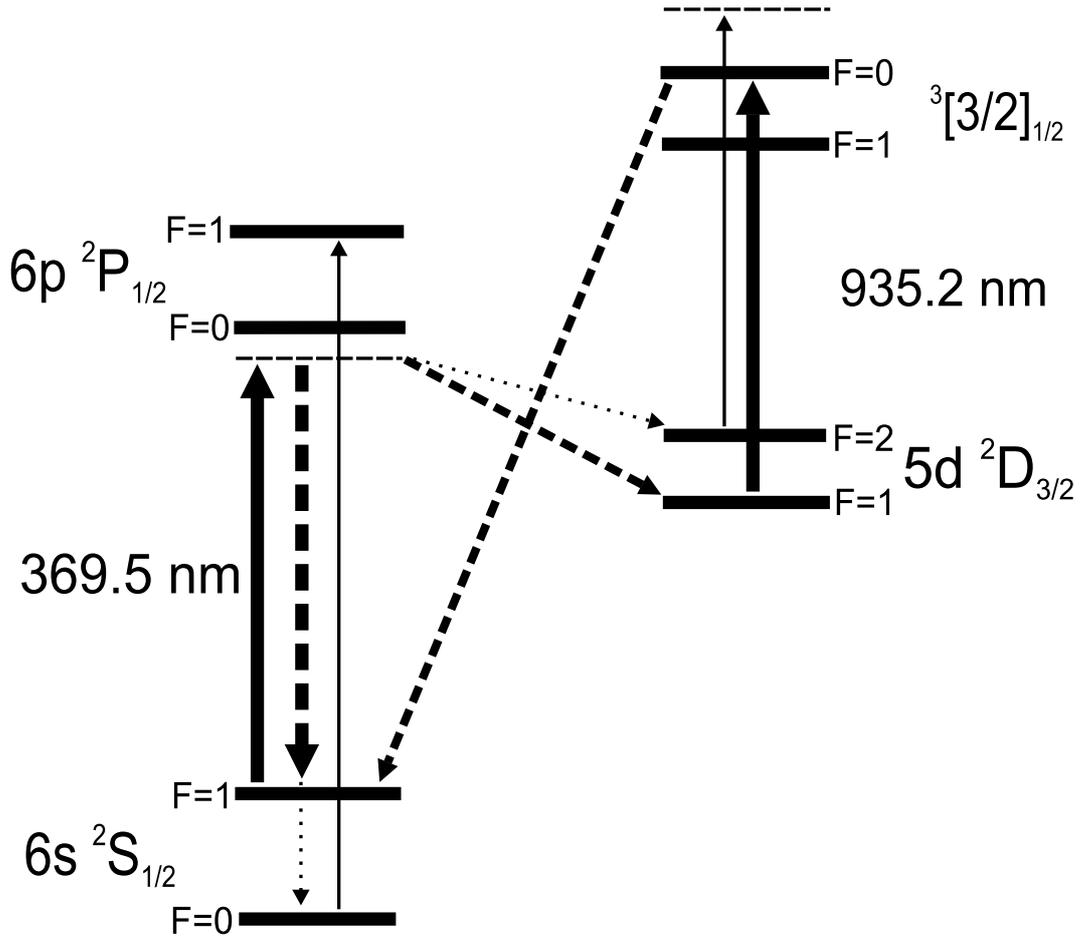

FIG. 1. Partial energy level system of $^{171}$Yb$^+$, showing the applied optical excitation (full arrows) and experimentally relevant spontaneous decay paths (broken arrows). The hyperfine splittings of the energy levels are not drawn to scale. The splitting frequencies are $\Delta_S \simeq 12.6$ GHz, $\Delta_P \simeq 2.1$ GHz, $\Delta_D \simeq 0.86$ GHz, $\Delta_{[3/2]} \simeq 2.5$ GHz.

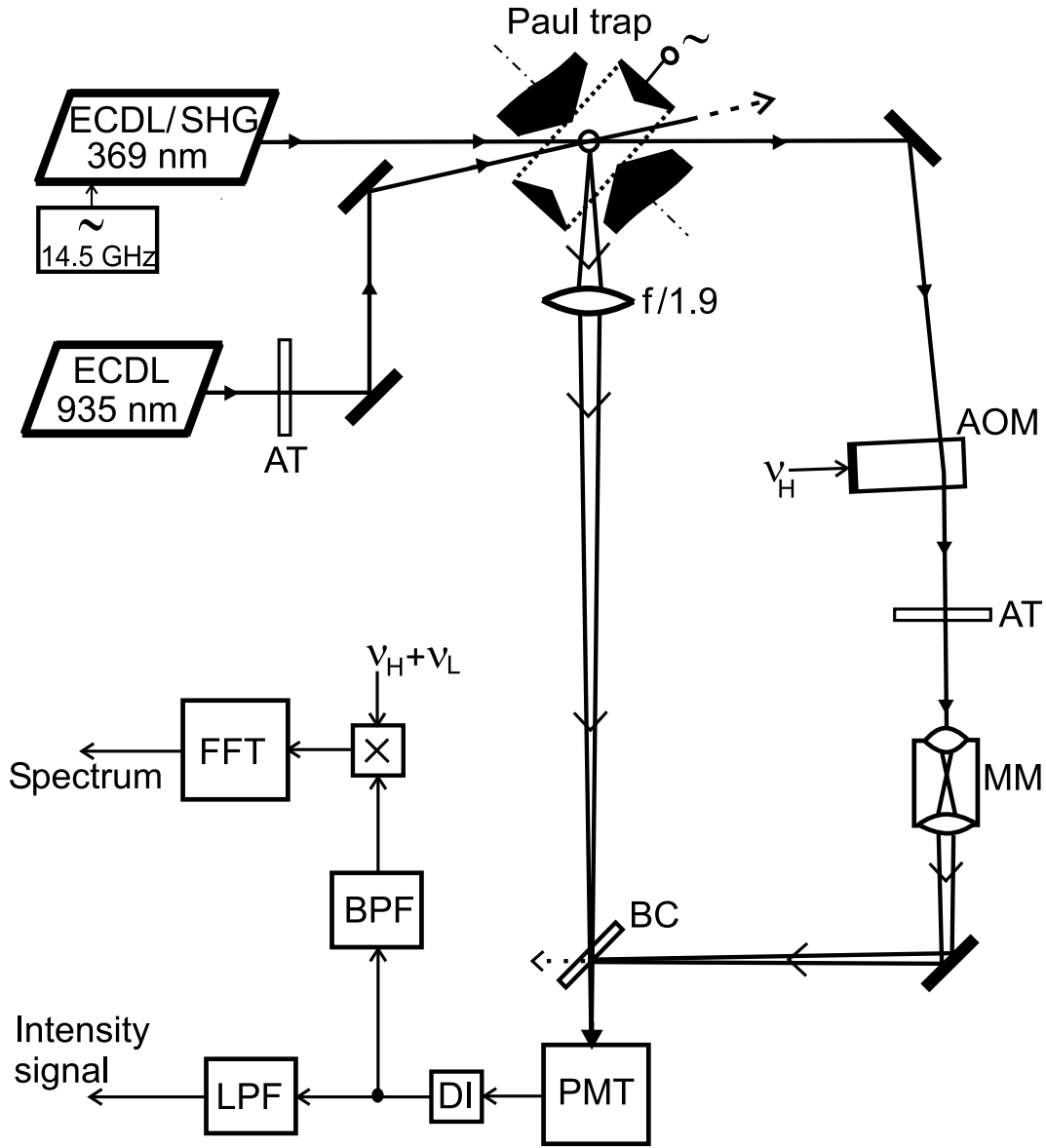

FIG. 2. Experimental scheme. ECDL, extended-cavity diode laser; SHG, second-harmonic generation; AT, attenuator; AOM, acousto-optic modulator; MM, mode-matching optic; BC, uncoated glass plate serving as beam combiner; PMT, photomultiplier tube; DI, pulse discriminator; LPF, low-pass filter; BPF, band-pass filter. The heterodyne frequency is $\nu_H = 21.4$ MHz. The heterodyne signal is down-converted to a frequency interval centered at $\nu_L = 70$ kHz. For further details see text.

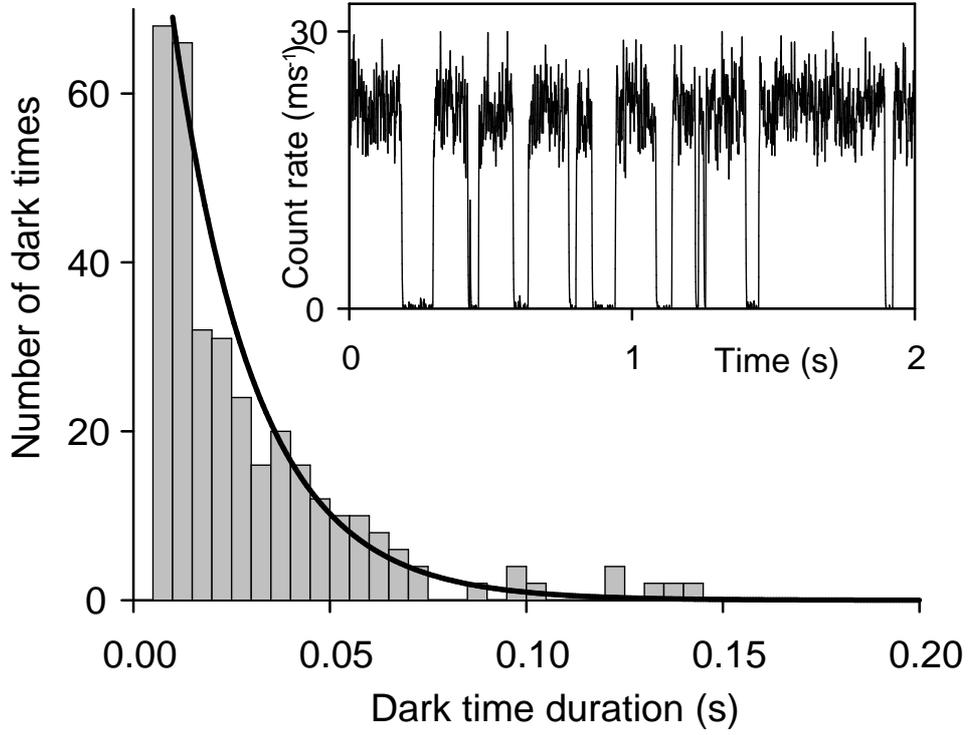

FIG. 3. Observed temporal variation of fluorescence intensity (inset) and histogram of dark-period durations for an applied 935-nm power of 0.4 mW. The data collection time was 60 s. The exponential fit to the histogram (solid line) indicates that the average dark-period length is $\tau_D = 21$ ms with an effective statistical uncertainty of $\pm 1.2$ ms. The number of dark periods with less than 5 ms duration was not determined since such dark periods are too short to be reliably resolved.

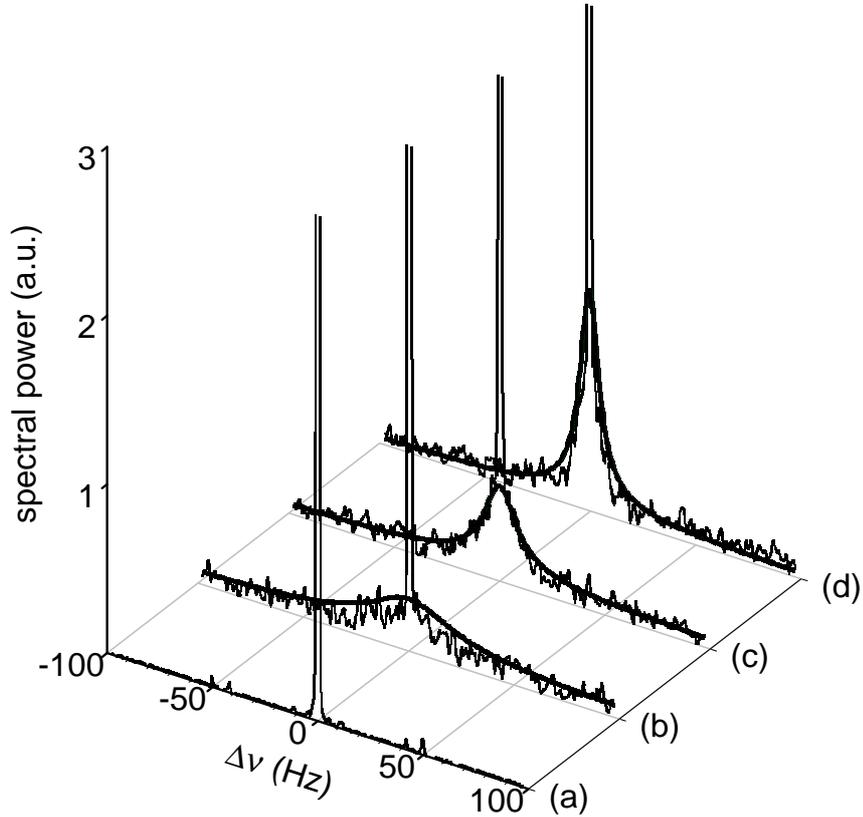

FIG. 4. Experimental fluorescence spectra of a trapped $^{172}$Yb ion (a), and of a $^{171}$Yb ion undergoing quantum jumps with measured average bright and dark period durations of $\tau_B = 103$ms, $\tau_D = 8$ms (b); $\tau_B = 171$ms, $\tau_D = 21$ms (c); $\tau_B = 160$ms, $\tau_D = 39$ms (d). The abscissa scale indicates the detuning from the optical excitation frequency at 370 nm. The smooth curves superimposed on the $^{171}$Yb$^+$ spectra are calculated Lorentzian pedestal lineshapes (see text). The observed full height of the Rayleigh peak at $\Delta\nu = 0$ is approximately 70 scale units in (a) and (b), and approximately 60 and 40 units respectively in (c) and (d). Broadband photodetection noise pedestals with heights of approximately 1 scale unit are suppressed in the displayed $^{171}$Yb$^+$ spectra.

14